\documentstyle[11pt]{article}

\catcode`@=11
\def\dif{{\rm d}}
\def\d{{\rm d}}
\def\deriv{\@ifnextchar[{\@deriv}{\@deriv[]}}
\def\@deriv[#1]#2#3{\mathchoice%
{{\dif^{#1}#2\over\dif{#3}^{#1}}}{{\dif^{#1}#2/\dif{#3}^{#1}}}%
{{\dif^{#1}#2\over\dif{#3}^{#1}}}{{\dif^{#1}#2/\dif{#3}^{#1}}}}
\def\derpar{\@ifnextchar[{\@derpar}{\@derpar[]}}
\def\@derpar[#1]#2#3{\mathchoice%
{{\partial^{#1}#2\over\partial{#3}^{#1}}}{{\partial^{#1}#2/\partial{#3}^{#1}}}%
{{\partial^{#1}#2\over\partial{#3}^{#1}}}{{\partial^{#1}#2/\partial{#3}^{#1}}}}

\def\secteqno{\@addtoreset{equation}{section}%
\def\theequation{\thesection.\arabic{equation}}}
\def\endsecteqno{\def\theequation{\@ifundefined{chapter}%
{\arabic{equation}}{\thechapter.\arabic{equation}}}}

\catcode`@=12
\secteqno

\def\beq{\begin{equation}}
\def\eeq{\end{equation}}

\newtheorem{rem}{Remark}

\newtheorem{theorem}{Theorem}

\newtheorem{prop}{Proposition}
\newtheorem{definition}{Definition}
\newtheorem{lemma}{Lemma}

\font\fr=eufm10  scaled \magstep 1  1

\def\df{{\mit\Omega}}
\def\vf{\mbox{\fr X}}
\def\inn{\mathop{i}\nolimits}
\def\Cinfty{{\rm C}^\infty}
\def\beann{\begin{eqnarray*}}
\def\eeann{\end{eqnarray*}}
\def\Lie{\mathop{\rm L}\nolimits}
\def\qed{\ifvmode\removelastskip\fi
{\unskip\nobreak\hfil\penalty50\hbox{}\nobreak\hfil
\hbox{\vrule height1.2ex width1.2ex}\parfillskip=0pt
\finalhyphendemerits=0 \par\smallskip}}
\def\dst{\(\displaystyle}
\def\proof{( {\sl Proof} )\quad}

\def\Real{{\bf R}}
\def\dif{{\rm d}}

\def\Tan{{\rm T}}

\begin{document}

\vskip 3mm

\begin{center}
{\large \bf GEOMETRIC REDUCTION IN OPTIMAL CONTROL THEORY WITH SYMMETRIES}
\\[7mm]
{\sc A. Echeverr\'{\i}a-Enr\'{\i}quez}\\[0.5mm] 
{\it Departamento de Matem\'atica Aplicada IV\\ 
Ed. C-3, Campus Norte UPC\\ C/. Jordi Girona 1-3. 08034 Barcelona, Spain}\\[4mm] 
{\sc J. Mar{\'\i}n-Solano}\\[0.5mm] 
{\it Departament de Matem\`atica Econ\`omica, Financera i Actuarial\\ 
Universitat de Barcelona\\
Av. Diagonal, 690. 08034 Barcelona, Spain\\[0.5mm] 
{\tt jmarin@eco.ub.es}}\\[4mm] 
{\sc M.C. Mu\~{n}oz-Lecanda}\\[0.5mm] 
{\it Departamento de Matem\'atica Aplicada IV\\ 
Ed. C-3, Campus Norte UPC\\ 
C/. Jordi Girona 1-3. 08034 Barcelona, Spain\\[0.5mm] 
{\tt matmcml@mat.upc.es}}\\[4mm] 
{\sc N. Rom\'an-Roy}\\[0.5mm] 
{\it Departamento de Matem\'atica Aplicada IV\\ 
Ed. C-3, Campus Norte UPC\\ 
C/. Jordi Girona 1-3. 08034 Barcelona, Spain\\[0.5mm] 
{\tt matnrr@mat.upc.es}} \vskip 4mm
\today
\end{center}

\pagestyle{myheadings}
\markright{A.\,Echeverr\'{\i}a-Enr\'{\i}quez \it et al: \quad
Geometric reduction in optimal control theory...}


\begin{abstract}
A general study of symmetries in optimal control theory is given,
starting from the presymplectic description of this kind of system.
Then, Noether's theorem, as well as the corresponding reduction procedure
(based on the application of the {\sl Marsden-Weinstein theorem}
adapted to the presymplectic case) are stated both in the
regular and singular cases, which are previously described.
\end{abstract}


\bigskip
\bigskip
\vbox{\raggedleft AMS s.\,c.\,(2000):
37J15, 49K15, 70G45, 70G65}\null

\newpage



\section{Introduction}

The application of modern differential geometry to
optimal control theory has meant a great advance for this field in recent years,
begining with R.W. Brocket's pioneering work
\cite{Br-70}, \cite{Br-73}, up to recent developments,
as reported, for instance, by H.J. Sussmann \cite{Su-95}, \cite{Su-98}, \cite{Su-99}.
This paper is devoted to studying
optimal control problems with symmetry in a geometric framework.

Hence, our standpoint is the natural presymplectic description of 
optimal control problems arising from the {\sl Pontryagin maximum principle}.
If this presymplectic system has symmetries, then the 
{\sl Marsden-Weinstein reduction theorem} \cite{MW-rsms}, 
generalized to the presymplectic case, as in \cite{EMR-99}, allows the dynamics
to be simplified, thus reducing the number of degrees of 
freedom and giving a simpler structure to the equations of motion.

Previous works on this subject have been carried out by
A.J. Van der Schaft \cite{Sh-87} and H.J. Sussmann \cite{Su-95}. The first
considers symmetries of the Lagrangian and the differential equation in order
to arrive at a Noether theorem (although no intrinsic geometric structures are used in it).
In the second, similar results are given, but in a more general context
(relaxing the differentiability conditions).
Recently, A. Bloch and P. Crouch offered a presentation of optimal
control systems on coadjoint orbits related to reduction
problems and integrability \cite{BC-99}.

In this work, we give a general description of symmetries in optimal control,
classifying them as the so-called {\sl natural} ones 
(which come from diffeomorphisms in the configuration
manifold of the original problem), and other symmetries of the associated
presymplectic system. Moreover, the reduction procedure is described both in the
regular and the singular case.
(A different point of view on this problem, using {\sl Dirac structures}
and {\sl implicit Hamiltonian systems} is given in
\cite{Bl-2002} and \cite{BVS-2000}. Other approaches can be found in
 \cite{Co-01}, \cite{DI-00} and  \cite{MCL-01}).

More precisely, Section 2 is devoted to stating the problem and describing the
presymplectic formulation. Since in this situation
there is no global dynamics, we study the application of a presymplectic
algorithm (see \cite{GNH-pca}) to obtain the (maximal) manifold where the dynamics exists
and the equation of motion on this last manifold. The analysis of the procedure leads
to a distinction between the {\sl regular} situation,
where only one step of the algorithm is needed, and the {\sl singular} one.
Later on, in Section 3, after
reviewing some basic facts of actions of Lie groups on
presymplectic manifolds, different notions of symmetries for
{\sl autonomous} control problems are
defined, and the reduction procedure as well as Noether's theorem are
stated for both the regular and singular cases.
In Section 4, the results are then extended to the case of
{\sl non-autonomous} control problems.
Finally, in Section 5, two examples are given:
In the first one the reduction of regular optimal
control problems invariant by a vector fields is considered,
following the study given in \cite{MR-01} for time-dependent
Lagrangian mechanics. The second corresponds to an example
analyzed by H.J. Sussman in \cite{Su-95}, and consists in
searching for the shortest paths with a bounded curvature.

All the manifolds are real, connected, second countable and
$C^\infty$. The maps are assumed to be $C^\infty$ and the
differential forms have constant rank. Sum over crossed repeated
indices is understood.

\section{Geometric description of optimal control theory}

\subsection{Optimal control problems}

Let $W=U\times V\subset\Real^n$ equipped with coordinates
$\{ q^i,u^a\}$ ($i=1,\ldots ,m$, $a=1,\ldots ,n-m$).
$\{ q^i\}$ are the coordinates in the {\sl configuration space}
$V\subset\Real^m$, and $\{ u^a\}$ are said to be the {\sl control variables}
or coordinates of the {\sl control space} $U\subset\Real^{n-m}$.
An {\sl optimal control problem} consists in finding $C^1$-piecewise
smooth curves $\gamma(t)=(q(t),u(t))$ with fixed endpoints in
configuration space, $q(t_1)=q_1$ and $q(t_2)=q_2$, such that they satisfy the
control equation
\begin{equation}\label{ligadura}
\dot{q}^i(t)=F^i(q(t),u(t))
\end{equation}
and minimize the objective functional
$$
S[\gamma]=\int_{t_1}^{t_2} L(q(t),u(t))\dif t\; .
$$
where $F^i,L\in\Cinfty (W)$.
Solutions to this problem are called {\sl optimal trajectories}
(relative to the points $q_1$ and $q_2$).

It is well-known (see \cite{Ju-97}) that the solution to this problem is provided by
{\sl Pontryagin's maximum principle}, which is stated in the following way:

First, consider the {\sl co-state space} $\Tan^*V$, whose coordinates are
denoted by $\{ q^i,p_i\}$ ($i=1,\ldots ,m$), and take $M\equiv U\times \Tan^*V$,
with coordinates $\{ q^i,p_i,u^a\}$. 
Then consider a family  $\{ H(q,p,u)\}\subset\Cinfty(M)$
of {\sl Hamiltonian functions}, parametrized by the control variables, given by
\begin{equation}\label{hamiltoniano}
H(q,p,u)=p_i F^i(q,u) - p_0 L(q,u)\; .
\end{equation}
where $p_0$ can be regarded as another parameter. 
For each control function $u(t)$, we can find the integral curves $(q(t),p(t))$ of
the {\sl Hamiltonian vector field} 
which are the solutions to the {\sl Hamilton equations}
\begin{equation}\label{eqham}
\dot{q}^i=\frac{\partial H(q,p,u)}{\partial p_i}\; ,~~~~~
\dot{p}^i=-\frac{\partial H(q,p,u)}{\partial q^i}\; 
\qquad (i=1,\ldots ,m).
\end{equation}
Secondly, the {\sl maximal Hamiltonian function} is defined as: 
\begin{equation}\label{Pon}
H_{max}(q,p)=\max_u H(q,p,u)\quad ,\quad \mbox{\rm for every $(q,p)\in\Tan^*V$}\; ,
\end{equation}
then we have:

{\bf Pontryagin's maximum principle}:
If a curve
$\gamma(t)=(q(t),u(t))$ is an optimal trajectory
between $q_1$ and $q_2$, then there exists a curve $p(t)$ such that:
\begin{enumerate}
\item
$(q(t),p(t),u(t))$ is the solution to (\ref{eqham}), and
\item
$H(q(t),p(t),u(t))=H_{max}(q(t),p(t))$.
\end{enumerate} 

If $p_0=0$, then the optimal solutions are
called {\sl abnormal}. In this paper we confine our attention to the
case in which $p_0\neq 0$, and in particular we take the typical value $p_0=1$.

It is clear that a necessary condition for $H$ to reach the maximum
(if the maximum of $H$ is not on the boundary of the control set) is
\begin{equation}\label{eqlig}
\chi_a=\frac{\partial H}{\partial u^a}=0\; ,\quad (a=1,\ldots ,n-m) .
\end{equation}
Hence, the trajectories solution to the optimal control problem
lie in a subset $M_1$ of the total space $M$,
which is defined by the constraints $\chi_a=0$.

In most cases, the constraint functions $\chi_a=0$, called 
{\sl first order constraints}, define implicitly $n-m$ functions $\psi^a$ such that
\begin{equation}\label{feedback}
u^a=\psi^a(q,p)
\end{equation}
whenever the matrix defined by
\dst W_{ab}=\frac{\partial\chi_a}{\partial u^b}\)
is non-singular, i.e., $\det W_{ab}\neq 0$. Under these circumstances, the function
$\psi\equiv\{ \psi^a\}$, given by equation (\ref{feedback}),
is called an {\sl optimal feedback function}. If such a condition is
satisfied, we say that the optimal control problem is {\sl regular}. 
If $\det W_{ab}=0$ on $M_1$, we say that the optimal control problem 
is {\sl singular}. In any case, we will assume that ${\rm rank}\,W_{ab}$
is constant on the domain of our analysis.

\begin{rem}
{\rm  One of the consequences of the Maximum Principle is that the
optimal control problems can be studied as {\sl presymplectic Hamiltonian systems};
that is, those where the 2-form is degenerate. Next we give a concise description
on this topic, as well as the associated constraint algorithm.
Furthermore, in Section \ref{tres}, we study the existence of first integrals
 for optimal control problems with symmetries (see Theorem \ref{propsym}).}
\end{rem}

\subsection{Optimal control problems as presymplectic Hamiltonian systems}

Taking into account the above considerations,
a problem of optimal control, from a geometric viewpoint, may be
given by the following data: a configuration space which is a differentiable
manifold $Q$, locally described by
the state variables $q^i$ ($i=1,\dots m$), a fibre bundle $\pi\colon E\to Q$ whose
fibres are locally described by the control variables $u^a$ ($a=1,\dots,n-m$),
a vector field $X$ along the projection of the bundle, $X\colon E\to\Tan Q$
(i.e., $\tau_Q\circ X=\pi$, where $\tau_Q\colon\Tan Q\to Q$ denotes the
canonical projection), and a ``Lagrangian function'' 
$L\colon E\to\Real$. Consider the family of paths
$\gamma\colon I\to E$ such that $\pi\circ\gamma$
has fixed end-points, which are solutions to the differential equation
\begin{equation}\label{difcon}
\Tan\pi\circ\dot{\gamma} = (\pi\circ\gamma)^{\textstyle.} =
X\circ\gamma
\end{equation}
that rules the evolution of the state variables, i.e., in local coordinates
it is equation (\ref{ligadura}), with boundary
conditions $q^i_1=q^i(t_1)$ to $q^i_2=q^i(t_2)$ (there are no
boundary conditions on the control variables).
The problem is to find a minimum of the
action
\[
\int_\gamma L(\gamma(t))\,\dif t
\]
for this family of paths $\gamma$. So we have the diagram
 \[
\begin{array}{ccccc}
 & & \Tan E &
\begin{picture}(30,10)(0,0) \put(13,6){\mbox{$\Tan\pi$}}
\put(0,3){\vector(1,0){40}}
\end{picture} & \Tan Q
\\
 &
 \begin{picture}(30,30)(0,0)
\put(8,19){\mbox{$\dot{\gamma}$}} \put(0,0){\vector(1,1){30}}
\end{picture}
&
\begin{picture}(10,30)(0,0)
\put(8,12){\mbox{$\tau_E$}} \put(3,30){\vector(0,-1){30}}
\end{picture}
&
\begin{picture}(30,30)(0,0)
\put(8,19){\mbox{$X$}} \put(0,0){\vector(1,1){30}}
\end{picture}
&
\begin{picture}(10,30)(0,0)
\put(8,12){\mbox{$\tau_Q$}} \put(3,30){\vector(0,-1){30}}
\end{picture}
\\
I &
\begin{picture}(30,10)(0,0)
\put(13,6){\mbox{$\gamma$}} \put(0,3){\vector(1,0){40}}
\end{picture}
& E &
\begin{picture}(30,10)(0,0)
\put(13,6){\mbox{$\pi$}} \put(0,3){\vector(1,0){40}}
\end{picture}
& Q
\end{array}
 \]
Therefore, an optimal control problem is characterized by the data $(E,\pi,Q,L,X)$.

\begin{rem}
{\rm It is easy to show that this is indeed a vakonomic problem on the
manifold~$E$ (see \cite{AR},\cite{GMM-00} for this kind of
problem), where the Lagrangian $L$ is singular, since it does
not depend on the velocities. The constraint submanifold
$C\subset\Tan E$, given by the differential equation above, is
$$
C=\{ w\in \Tan E\mid\Tan\pi (w)=(X\circ\tau_E)(w)\}\; .
$$
In this way, a path $\gamma$ is admissible if, and only if, it is
a solution to the differential equation (\ref{difcon}) or,
equivalently, if it takes values in the affine subbundle $C$ of
$\Tan E$. Notice that, in coordinates, and from a
variational viewpoint, the constraints defining $C$ as a submanifold of $\Tan E$
(which are the set of first order differential equations
(\ref{ligadura})) are very particular: they
give the velocities of the state variables in terms of the
state and control variables}.
\end{rem}

Optimal control theory admits several geometric formulations and
expressions of the equations of motion (\ref{eqham}) and
(\ref{eqlig}). One of the most interesting is the {\it presymplectic
description}\/ which can be constructed on the manifold
$\pi^*\Tan^*Q=E\times_Q\Tan^*Q$
(for a description on presymplectic dynamical systems, in general, see
for instance \cite{Ca-90}, \cite{GNH-pca}). We denote by $\pi_1\colon
\pi^*\Tan^*Q \to E$ and $\pi_2\colon \pi^*\Tan^*Q \to \Tan^*Q$ the
projections onto the first and second factors, respectively. Then,
the Hamiltonian function (\ref{hamiltoniano}) can be defined
intrinsically as
 \begin{equation}\label{hamgeom}
 H = \hat{X} - L\; ,
 \end{equation}
 where we identify the Lagrangian function $L\in\Cinfty(E)$ with
 its pull-back through $\pi_1\colon\pi^*\Tan^*Q\to E$, and
 $\hat{X}\colon\pi^*\Tan^*Q\to\Real$ is the function defined by
 $\hat{X}((q,u),\alpha)=\alpha(X(q,u))$, for $(q,u)\in E$ and $\alpha\in\Tan_q^*E$.

Let $\theta_0$ and $\omega_0=-\dif\theta_0$ be the canonical 1 and
2-forms in $\Tan^*Q$. We can take their pull-back through
$\pi_2$, obtaining $\theta=\pi_2^*\theta_0$ and $\omega=\pi_2^*\omega_0$
in $\pi^*\Tan^*Q$. In local coordinates, $\theta=p_i\dif q^i$ and
$\omega=\dif q^i \wedge \dif p_i$.  If $\eta$ is a path on
$\pi^*\Tan^*Q$, then to solve the presymplectic equation
$$
\inn_{\dot\eta} \omega = \dif H \circ \eta
$$
is equivalent to solving the equations of motion (\ref{eqham}) and (\ref{eqlig}). To
show this, it is enough to write their local expressions. If $\eta$
is an integral curve of a vector field $\Gamma$ on $\pi^*\Tan^*Q$,
we can write the presymplectic equation above as
\begin{equation}\label{conmo}
\inn_\Gamma\omega=\dif H\; .
\end{equation}
Summarizing, our geometrical interpretation of 
the problem of optimal control stated in the begining is
given by the presymplectic Hamiltonian system
$(\pi^*\Tan^*Q,\omega, H)$, which arises from the original data $(E,\pi,Q,L,X)$.

Observe that relation (\ref{conmo}) gives the critical curves of
the variational problem for the Lagrangian $L$ with constraints
(\ref{difcon}), and it is a weaker condition, in general, than the
maximum principle posed by equations (\ref{eqham}), (\ref{Pon}).
Througout this paper, we
will restrict our attention to the analysis of solutions to
(\ref{conmo}), and the existence of true extremals will not be studied.

 From now on, we will denote $M=\pi^*\Tan^*Q$. Relation
(\ref{eqlig}) is the local expression of the compatibility condition for the equation
(\ref{conmo}) and it is assumed that it defines a closed submanifold $M_1$ in $M$.
In the following section, all these features will be studied in detail, and
we show that the regularity of the optimal control problem is
equivalent to the existence of a unique vector field solution to
(\ref{conmo}) tangent to the first order constraint submanifold
$M_1$. Otherwise, the optimal control problem is singular, and a
constraint algorithm is needed to solve the problem, in general.

\subsection{Constraint algorithm for optimal control problems}

Consider the presymplectic dynamical system given by  $(M,\omega,H)$, with
$M=\pi^*\Tan^*Q$, $\omega=\pi^*_2\omega_0$, and $H$ being defined by
(\ref{hamgeom}). The presymplectic dynamical equation is (\ref{conmo}).
Notice that $\omega$ is degenerate, and
$\ker\omega=\vf^{{\rm V}(\pi_2)}(M)=\{ X\in\vf (M)\,\mid\, \pi_{2*}X=0\}$. 
It is well known \cite{GNH-pca}, \cite{MR-92} that there are vector fields
satisfying equation (\ref{conmo}) only at the points of the subset
$$
M_1=\{ x\in M\, \mid\, (\Lie_ZH)(x)=0\, ,\ \mbox{for every $Z\in\ker\omega$}\}
$$
We assume that $M_1$
is a closed submanifold of $M$, and the natural embedding is denoted by
$j_1\colon M_1\hookrightarrow M$.
Now, there are vector fields  $\Gamma$ satisfying equation (\ref{conmo})
on the points of $M_1$. These vector fields are defined in principle only at
the points of $M_1$, and they take values in $\Tan M|_{M_1}$
(obviously, they can be extended to vector fields in $M$).
However, in general, these vector fields $\Gamma$ are not tangent to $M_1$,
that is, they do not take values in $\Tan M_1$.
If ${\rm ker}\,\omega\cap\underline{\vf(M_1)}=\{ 0\}$
(where $\underline{\vf(M_1)}$ denotes the vector fields of $M$
which are tangent to $M_1$), then $(M_1,j_1^*\omega)$
is a symplectic manifold, and there is a unique vector field $\Gamma$
defined at the points of $M_1$ verifying (\ref{conmo}).
Moreover $\Gamma$ is tangent to $M_1$ because $j_1^*\omega$ is a symplectic form.
As locally $\ker\omega =\langle \partial / \partial u^a\rangle$,
then $M_1$ is a submanifold transverse to $\ker\omega$ if, and only if,
$\det W_{ab}\neq 0$. So, a regular optimal control problem corresponds to the case
where $(M_1,j_1^*\omega)$ is a symplectic manifold.

However, if the system is singular, the vector fields solutions to
the Hamiltonian equation (\ref{conmo})
on the submanifold $M_1$ are not necessarily tangent to $M_1$.
Thus, their integral curves can leave the submanifold
where the extremal trajectories must lie. Thus, we must take the points
of $M_1$ where vector fields solutions to (\ref{conmo})
being tangent to $M_1$ exist. The subset $M_2\subset M_1$ made by those points
is defined as
\[
M_2=\{ x\in M_1\, \mid\, \Gamma(\chi_a)(x)=0\, , 
\mbox{ for every $\Gamma$ solution to (\ref{conmo}) on $M_1$}\}\; .
\]
We assume that the
subset $M_2$ is a closed submanifold of $M_1$. We denote the
functions defining $M_2$ on $M_1$ by $\chi^{(2)}_b$. Repeating the
argument, we obtain a family of subsets (assuming that all of
them are closed submanifolds) defined recursively by
\[
M_k=\{ x\in M_{k-1}\, \mid\, \Gamma(\chi^{(k-1)}_b)(q)=0,
\mbox{ for every $\Gamma$ solution to (\ref{conmo}) on $M_1$}\}\; ,~~k>1\; .
\]
The recursion stops, and $M_r=M_{r+1}=M_{r+2}=\cdots$
for a certain $r$. In this way, we obtain a stable
submanifold
\[
M_f=\cap_{k\geq 1} M_k
\]
where the dynamical equation has tangent solutions, and the integral curves
of the corresponding vector fields are the critical curves
of the singular optimal control problem. We denote by
$j_f\colon M_f\hookrightarrow M$ the natural embedding.
This is the {\sl constraint algorithm for optimal control problems},
which is similar in nonlinear control to the so-called {\sl zero-dynamics algorithm}
based on the notion of {\sl (locally) controlled invariant submanifold}
(see \cite{Is-95}, \cite{NS-90}).

Another geometric description of the
condition of regularity and the submanifold $M_1$ can be given.
In fact, as we know that $M=\pi^*\Tan^*Q=E\times_Q\Tan^*Q$, we can  consider
the fibre bundle $\pi_2\colon E\times_Q\Tan^*Q\to\Tan^*Q$.
Now, consider the
Hamiltonian function $H\colon E\times_Q\Tan^*Q\to \Real$, and the
vertical bundle $V(\pi_2)$. The function $H$ defines a map
 \[
 \begin{array}{ccc}
 {\cal F}H\colon E\times_Q\Tan^*Q & \longrightarrow &
V^*(\pi_2) \\ (e,\alpha) & \mapsto & \Tan_e\left(
H|_{(\pi(e),\alpha)}\right) \colon \mbox{V}(\pi_2)\to\Real
 \end{array} \]
which is called the {\sl fibre derivative} of $H$ (see \cite{Gr00},\cite{Go73}
for more details). In local coordinates, 
${\cal F}H(q,p,u)=(q,p,u,\partial H/\partial u)$.
Then, the submanifold
$M_1$ can be characterized as $M_1={\cal F}H^{-1}(0)$. 
The local regularity
condition $\det W_{ab}\neq 0$ is equivalent to demanding that 
${\cal F}H$ has maximal rank everywhere, and it is also
equivalent to the existence of a
(local) section $\sigma\colon U\to E\times_Q\Tan^*Q$, for some
neighbourhood $U$ of each point $(q,p)\in\Tan^*Q$.
If there exists a global section $\sigma$ of $\pi_2$, such that 
${\cal F}H^{-1}(0)=\sigma(\Tan^*Q)$, then the optimal control problem is
said to be {\sl hyper-regular} and, of course, $\det W_{ab}\neq 0$
(i.e., it is regular).

In the regular case, the 2-form $\omega_1=j_1^*\omega$ is non-degenerate
and the manifold $M_1$ is locally symplectomorphic to $\Tan^*Q$.
In the hyper-regular situation, $M_1$ and $\Tan^*Q$ are globally
symplectomorphic, and the symplectomorphism is constructed by
means of the global section $\sigma$.
As a final remark, if $\Gamma_1\in\vf(M_1)$ is the unique vector field
solution to the dynamical equation
 \begin{equation}\label{conmor}
 \inn_{\Gamma_1}\omega_1=\dif h_1\; ,
 \end{equation}
 where $h_1=j_1^*H$, then the vector field $\Gamma$ solution to
 (\ref{conmo}) is obtained from $\Gamma_1$
 by using the optimal feedback condition (\ref{feedback}).

Notice that, in both the regular and singular cases, there is
no vector field on $M$ satisfying the presymplectic dynamical
equation in $M$, but only on a submanifold $M_1\not= M$.

\section{Symmetries and reduction of optimal control problems:
 the autonomous case}
\protect\label{tres}

(See the appendix \ref{algpm} for the notation, the terminology and the
fundamental concepts and results about presymplectic reduction,
which are used in this section).

\subsection{The regular case: symmetries and reduction}

\subsubsection{Symmetries and first integrals}

One of the most important features in the study of dynamical
systems with symmetry is the so-called {\it reduction theory}.

First, we establish the concept of group of symmetries for the
non-compatible presymplectic dynamical system $(M,\omega ,H)$,
where $M=\pi^*\Tan^*Q$ and $H$ is given by (\ref{hamgeom}), with
compatible symplectic dynamical system $(M_1,\omega_1,h_1)$, where
$\omega_1=j_1^*\omega$ and $h_1=j_1^* H$; i.e., we assume that the
optimal control problem is regular. Notice that, in this case, in
the notation of the above section, $(P,\Omega)=(M_1,\omega_1)$.
Moreover, $\omega_1$ is symplectic and exact, because
$\omega_1=j_1^*\omega = j_1^*(-\dif\theta) = -\dif(j_1^*\theta)$.

\begin{definition}\label{sym1} Let $G$ be a connected Lie group and
$\Phi\colon G\times M\to M$ an action of $G$ on $M$. Let
$(M,\omega,H)$ be a regular optimal control problem. $G$ is said
to be a symmetry group of $(M_1,\omega_1,h_1)$ if
\begin{enumerate}
\item
$\Phi$ leaves $M_1$ invariant; that is, it induces an action
$\Phi_1 \colon  G\times M_1\to M_1$.
\item
The induced action $\Phi_1$ is a symplectic action on
$(M_1,\omega_1 )$ (which is assumed to be Poissonian, free and
proper); that is, for every $g\in G$, $(\Phi_1)_g^*\omega_1 =
-\omega_1$.
\item
For every $g\in G$, $(\Phi_1)_g^* h_1=h_1$.
\end{enumerate}
\end{definition}

This definition is justified since, if $G$ is a symmetry group of
$(M_1,\omega_1,h_1)$, then $\Phi_g$ maps solutions in solutions.
To show this, let $\Gamma$ be the vector field in $M$ tangent to
$M_1$ solution to the dynamical system (\ref{conmo}) in the points
of $M_1$. Then there is a vector field $\Gamma_1\in\vf (M_1)$ such
that $j_{1*}\Gamma_1=\Gamma|_{M_1}$ and verifying
$\inn_{\Gamma_1}\omega_1=\dif h_1$. Therefore,
\[
0=(\Phi_1)_g^*(\inn_{\Gamma_1}\omega_1-\dif h_1) =
\inn_{(\Phi_1)_g^*\Gamma_1}(\Phi_1)_g^*\omega_1 - (\Phi_1)_g^*\dif
h_1 =
\]
\[
= \inn_{(\Phi_1)_g^*\Gamma_1}\omega_1 - \dif(\Phi_1)_g^* h_1 =
\inn_{(\Phi_1)_g^*\Gamma_1}\omega_1 - \dif h_1\; .
\]

This definition of symmetry group applies to every presymplectic
dynamical system, simply by identifying $M_1$ with the final
constraint submanifold $M_f$, in the sense that it works for the
case when the 2-form $\omega_1$ (or $\omega_f$ in the general
case) is degenerate (in that case, in Condition 2, the action
$\Phi_f$ is a presymplectic action on $(M_f,\omega_f)$).

In Definition \ref{sym1}, we have considered actions on $M$ which
induce symmetries on the symplectic manifold $M_1$ if the optimal
control problem is regular (symmetries on the presymplectic final
constraint submanifold, in general, if the problem is singular).
It is straightforward to show that if $G$ is a symmetry group of
$(M,\omega,H)$ (i.e., $\Phi_g^*\omega = \Omega$ and $\Phi_g^*H=H$,
for all $g\in G$), then it is a symmetry group of the associated
compatible presymplectic dynamical system. However, the group of
symmetries of $(M,\omega,H)$ is smaller, in general, than the
group of symmetries of $(M_f,\omega_f,h_h)$, in the sense that it
gives fewer symmetries of the dynamics $\Gamma_f$.

There is a more natural definition of transformation of symmetry
for optimal control systems strongly related to the special
characteristics of the problem. Let us recall that an optimal
control problem may be given by the data $(E,\pi,Q,X,L)$, where
$Q$ is the configuration space describing the state variables,
$\pi\colon E\to Q$ is a fibre bundle whose fibres describe the
control variables, $X\colon E\to\Tan Q$ is a vector field along
$\pi$ (i.e., $\tau_Q\circ X=\pi$), and $L\colon E\to\Real$ is a
Lagrangian function.

\begin{definition} Let $\Psi\colon E\to E$ be a bundle diffeomorphism,
and $\varphi\colon Q\to Q$ the induced diffeomorphism on the base
manifold ($\pi\circ\Psi=\varphi\circ\pi$). We say that $\Psi$ is a
transformation of symmetry of the regular optimal control problem
described by the data $(E,\pi,Q,X,L)$ if
\begin{enumerate}
\item $\Psi^* L=L$ $(\Psi$ is a symmetry of the Lagrangian
function$)$.
\item $\Psi_* X=X$ $(\Psi$ is a symmetry of the vector field$)$.
\end{enumerate}
\end{definition}
In the definition, the push-forward $\Psi_*X$ is defined by
 \begin{equation}\label{push}
 \Psi_*X=\varphi_*\circ X\circ\Psi^{-1}\; .
 \end{equation}
 Hence the inverse $\Psi^*$ of the push-forward is
 $\Psi^*X=\varphi_*^{-1}\circ X\circ\Psi$.

If the diffeomorphism $\Psi$ is understood as a change of coordinates
in the state and control variables then,
in nonlinear control, it is called as
{\sl feedback transformation}, and the push-forward
$\Psi_*X$ is the differential equation $X$ transformed via the feedback transformation
(see \cite{Is-95}, \cite{Ja-90}, \cite{NS-90}).
Moreover, $\Psi$ being a symmetry of $X$ means that the induced diffeomorphism
$\varphi$ is a symmetry of $X$ in the sense of \cite{Ja-98} and \cite{RT-2002}.

The meaning of this definition will become clear in Theorem
\ref{propsym}, where we prove that, if $G$ is a connected Lie
group such that $\Psi_g$ is a transformation of symmetry of the
regular optimal control problem, for every $g\in G$, then $\Psi_g$
maps optimal trajectories into optimal trajectories (see also
\cite{DI-00}).

Now, let $Z\in\vf(E)$ be a vector field on $E$, and let $X\colon
E\to \Tan Q$ be a vector field along the projection $\pi\colon
E\to Q$. If $Z$ is $\pi$-projectable, then we can define the Lie
derivative of $X$ along $Z$ as follows: if $\Psi_t$ denotes the
flow of $Z$ and $\varphi_t$ denotes the flow of the vector field
$Z_0=\pi_*Z\in\vf(Q)$, then
 \begin{equation}\label {push1}
 \Lie_ZX=\left.\frac{\dif}{\dif t}\right|_{t=0}\left[ \Psi_t^*
 X)\right]\; ,
 \end{equation}
 or, equivalently,
  \begin{equation}\label{push2}
 \Lie_ZX=\lim_{t\to 0}\frac{\Psi^*_tX-X}{t}\; .
 \end{equation}
 Notice that the push-forward is well-defined since from the
 projectability of $Z$ we deduce that
 $\Psi_t$ is a bundle mapping. It is clear that
 $\Lie_ZX\colon E\to\Tan Q$ is a vector field along $\pi$.

 The following lemma gives the algebraic expression of this Lie
 derivative.

 \begin{lemma}\label{Lie-der}
 Let $X\colon E\to \Tan Q$ be a vector field along the projection
 $\pi\colon E\to Q$, and let $Z\in\vf(E)$ be a $\pi$-projectable
 vector field on $E$, with $Z_0=\pi_*Z\in\vf(Q)$. Then, for
 every $f\in\Cinfty(Q)$, the Lie derivative
 $\Lie_ZX(f)\in\Cinfty(E)$ is
 \[ \Lie_ZX(f)= \Lie_Z (\Lie_X(f)) - \Lie_X(\Lie_{Z_0}(f))\; , \]
 i.e., $\Lie_Z(X)=\Lie_Z\circ\Lie_X - \Lie_X\circ\Lie_{Z_0}$.
 \end{lemma}

 \proof From (\ref{push2}) we must evaluate
 $(\Psi_t^*X-X)(f)(m)$, for every $f\in\Cinfty(Q)$ and $m\in E$.
 Using (\ref{push}), we obtain
 \begin{eqnarray}
 (\Psi^*_tX-X)(f)(m) & = & X_{\Psi_t(m)}(f\circ\varphi_{-t}) - X_m(f)
 = \nonumber \\
 & = & X_{\Psi_t(m)}(f\circ\varphi_{-t}) - X_{\Psi_t(m)}(f) +
 X_{\Psi_t(m)}(f)- X_m(f)\; . \nonumber
 \end{eqnarray}
 Hence,
 \[
 (\Lie_ZX)_m(f) = \lim_{t\to
 0}\left(\frac{\Psi_t^*X-X}{t}\right)_m(f)
 = \lim_{t\to
 0}\frac{X_{\Psi_t(m)}(f)-X_m(f)}{t} + \]
 \[ + \lim_{t\to
 0}\frac{X_{\Psi_t(m)}(f\circ\varphi_{-t})-X_{\Psi_t(m)}(f)}{t}
 =
 (\Lie_Z(\Lie_X(f)))_m - (\Lie_X(\Lie_{Z_0}(f)))_m \; .
 \]
 \hfill $\Box$

\begin{definition}\label{infisym} Let $Z\in\vf(E)$ be a
$\pi$-projectable vector field. The vector field $Z$ is called an
{\sl infinitesimal symmetry} of the regular optimal control
problem $(L,\pi,Q,X,L)$ if
\begin{enumerate}
\item $\Lie_ZL=0$ $(Z$ is an infinitesimal symmetry of the Lagrangian
function$)$.
\item $\Lie_ZX=0$ $(Z$ is an infinitesimal symmetry of the
vector field $X)$.
\end{enumerate}
\end{definition}

We finish these definitions of symmetries with the idea of
symmetry group of an optimal control problem.

\begin{definition} Let $G$ be a connected Lie group, and
$\Psi\colon G\times E\to E$ an action of $G$ on $E$ such that, for
each $g\in G$, $\Psi_g$ is a bundle mapping, with induced mapping
$\varphi_g\colon Q\to Q$. $G$ is said to be a {\sl symmetry group}
of the regular optimal control problem described by
$(E,\pi,Q,X,L)$ if every $\Psi_g$, $g\in G$, is a transformation
of symmetry.
\end{definition}

This concept of symmetry group of regular optimal control problems
is related to the idea of symmetry group of the presymplectic
dynamical system $(M,\omega, H)$ (or, equivalently, symmetry group
of $(M_1,\omega_1,h_1)$, since the problem is regular) as follows:
given the above action $\Psi\colon G\times E\to E$ preserving the
bundle structure $\pi\colon E\to Q$, we can lift this action to an
action $\Phi\colon G\times \pi^*\Tan^*Q\to \pi^*\Tan^*Q$ in a
natural way: for every $(q,u,p)\in \pi^*\Tan^*Q$ (where $u\in E_q$
and $p\in\Tan_q^*Q$),
\begin{equation}\label{actionlift}
\Phi_g(q,u,p)=\left( \Psi_g(q,u),\Tan^*_{\varphi_g
(q)}\varphi_g^{-1}(p)\right)\; .
\end{equation}

\begin{theorem}\label{propsym} If $G$ is a symmetry group of
the regular optimal control problem $(E,\pi,Q,X,L)$ (i.e.,
$\Psi_g\colon E\to E$ is a transformation of symmetry, for every
$g\in G$), then the action $\Phi$ given by (\ref{actionlift}) is a
symmetry group of the presymplectic dynamical system
$(\pi^*\Tan^*Q,\omega,H)$. Moreover, the action is exact and there
exists a comomentum map
\[ \begin{array}{ccc}
{\cal J}^*\colon{\bf g} & \longrightarrow & C^\infty(\pi^*\Tan^*Q)
\\ \xi & \mapsto & \inn_{\tilde{\xi}}\theta
\end{array} \]
in such a way that the functions $f_\xi=\inn_{\tilde{\xi}}\theta$
are constants of motion.

Conversely, if the lifted transformations $\Phi_g$ are symmetries
of the presymplectic dynamical system, then the fundamental vector
fields $\tilde{\xi}$ are infinitesimal symmetries of
$(E,\pi,Q,X,L)$.
\end{theorem}

\proof On the one hand, the action $\Psi$ is a bundle mapping (for
every $g\in G$, $\varphi_g$ is an action on the base manifold
$Q$), in such a way that $\varphi^*_g\colon\Tan^* Q\to\Tan^*Q$ is
symplectic and exact ($\varphi_g^*\theta_0=\theta_0$). Then, since
the new action $\Phi$ is the canonical lift of $\Psi$ to an action
in $\pi^*\Tan^*Q$ and $\theta=\pi_2^*\theta_0$, then
$\Phi_g^*\theta=\theta$ ($\Phi$ is exact), so we have
$\Phi_g^*\omega=\omega$. On the other hand, $\Phi_g^*H
=
\Phi_g^*\hat{X} -\Phi_g^*L = \hat{X} - L=H$. Finally, the
existence of the comomentum map follows from the exactness of the
action.

Conversely, notice that if $\Psi\colon G\times E\to E$
is a bundle action such that $G$ is a symmetry group of the
presymplectic system $(\pi^*\Tan^*Q,\omega,H)$, then $\Phi^*(H)=H$
implies that $\tilde{\xi}^c(H)=0$, where $\tilde{\xi}^c$ are the
fundamental vector fields associated with the action $\Phi$. It is
clear that $\tilde{\xi}^c$ are the lifting to $\pi^*\Tan^*Q$ of
the fundamental vector fields $\tilde{\xi}$ associated with the
action $\Psi$. The local expressions of such fundamental vector
fields associated with actions $\Psi$ and $\Phi$ are
\beann
\tilde{\xi} &=&
\xi^i(q)\frac{\partial}{\partial q^i}
+\zeta^a(q,u)\frac{\partial}{\partial u^a}
\\ 
 \tilde{\xi}^c &=& \xi^i(q)\frac{\partial}{\partial q^i} -
p_i\frac{\partial\xi^i(q)}{\partial q^j}\frac{\partial}{\partial
p_j}+\zeta^a(q,u)\frac{\partial}{\partial u^a}\; .
\eeann
 Then,
\beann
\tilde{\xi}^c(H)&=&
\tilde{\xi}^c\left(
p_iX^i(q,u)\right)-\tilde{\xi}^c\left( L(q,u)\right) =
\\ &=&
 p_i\left(\xi^j\frac{\partial X^i}{\partial q^j} -
Y^j\frac{\partial\xi^i}{\partial q^j}+\zeta^a\frac{\partial
Y^i}{\partial u^a}\right) - \left(\xi^i\frac{\partial L}{\partial
x^i} + \zeta^a\frac{\partial L}{\partial u^a}\right) = 0\; .
\eeann
Therefore, as $p_i$ are free, we obtain
\begin{eqnarray}\label{uno}
\xi^j\frac{\partial Y^i}{\partial x^j} -
Y^j\frac{\partial\xi^i}{\partial x^j}+\zeta^\alpha\frac{\partial
Y^i}{\partial u^\alpha} & = & 0\; ,~~~\hbox{and} \\ \label{dos}
\xi^i\frac{\partial L}{\partial x^i} + \zeta^\alpha\frac{\partial
L}{\partial u^\alpha} & = & 0\; .
\end{eqnarray}

 But, from Lemma \ref{Lie-der}, equation (\ref{uno}) is the expression
 in local coordinates of Condition 2 in Definition \ref{infisym},
$\Lie_{\tilde{\xi}^c}X=0$, and equation (\ref{dos}) means
$\tilde{\xi}(L)=0$, which is is equivalent to Condition 1 in the
same definition. Therefore, the fundamental vector fields
$\tilde{\xi}$ are infinitesimal symmetries of the optimal control
problem $(E,\pi,Q,X,L)$.
 \hfill $\Box$

We can compare this statement with other versions of Noether's theorem in control theory.
For instance, in \cite{Ju-95}, a particular optimal control problem on
a Lie group is studied: the conserved quantity related to the Casimir of the Lie group
is used to find the shortest path of a car moving under the suitable conditions.
In \cite{Sh-87}, the author studies the reduction of the Hamiltonian system
associated with an optimal control problem (by the Maximum Principle),
by a local symmetry given by a vector field.
In \cite{Su-95}, given an optimal control problem,
the author uses the Maximum Principle to construct a
family of symplectic Hamiltonian problems parametrized by the controls 
(instead of using the presymplectic alternative).
Then, constants of motion are related to the action of a
Lie algebra on $M$. The techniques are symplectic but local, and
reduction is not studied.
Finally, the situation studied in Theorem 3 in \cite{To-2002} is the following:
if an optimal control problem, formulated on $\Real^n$,
 is invariant by a one-parameter family of $C^1$-maps,
then a conservation law is obtained.

\begin{rem}
 {\rm As mentioned in Section 2, an optimal control
problem can be understood as a vakonomic problem where the
Lagrangian function $L\colon \Tan E\to\Real$ is a basic function and
the constraint submanifold $C$ is the affine subbundle locally
described by the contraints $\dot{q}^i=X^i(q,u)$, $i=1,\dots,n$.
Following Arnold {\sl et al} \cite{AR}, a transformation of
symmetry of the vakonomic system is a diffeomorphism $\Phi\colon
\Tan E\to\Tan E$ such that $\Phi|_C(C)\subset C$ and
$\Phi^*(L|_C)=L|_C$. But $L|_C=L$, since $L$ is a basic function,
so this last condition can be written as $\Phi^*(L)=L$ in
this case. It is easy to show that these two conditions are the
conditions we have assumed above when $\Phi$ is a diffeomorphism
adapted to the bundle structure.}
\end{rem}

\subsubsection{Momentum map and geometric reduction}

Now, if we have the compatible dynamical system $(M_1,\omega_1,h)$
and the action $\Phi_1$, we are interested in removing the
symmetries by following a reduction procedure in order to get a
symplectic dynamical system. We apply the results of the appendix
\ref{algpm}, where now $P\equiv M_1$ and $\Omega\equiv \omega_1$
is symplectic and exact. In what follows we assume that the action is
Poissonian, free and proper.

Let ${\cal J}$ be the momentum map associated with this action,
$\mu\in{\bf g}^*$ a weakly regular value, 
$\jmath_\mu\colon{\cal J}^{-1}(\mu )\hookrightarrow M$
the natural imbedding, and $\omega_\mu =j_\mu^*\omega_1$ and
 $h_\mu =j_\mu^*h_1$. Therefore:

\begin{prop}
$({\cal J}^{-1}(\mu ),\omega_\mu ,h_\mu)$
is a compatible presymplectic Hamiltonian system.
\end{prop}
\proof
 If we denote by ${\bf \tilde{g}_{M_1}}$ the set of
fundamental vector fields on $M_1$ with respect to the action
$\Phi_1$, and $\Gamma_1$ is the Hamiltonian vector field
associated with the Hamiltonian function $h_1$, then, for every
constraint $\zeta$, with $\d\zeta
=\inn_{\tilde{\xi}^{M_1}}\omega_1$,
$\tilde{\xi}^{M_1}\in\tilde{\bf g}_{M_1}$, defining ${\cal
J}^{-1}(\mu )$, $$ j_\mu^*\Gamma_1(\zeta )=
j_\mu^*(\inn_{\Gamma_1}\d\zeta )= j_\mu^*(\inn_{\Gamma_1}
\inn_{\tilde{\xi}^{M_1}}\omega )=
-j_\mu^*(\inn_{\tilde{\xi}^{M_1}} \inn_{\Gamma_1}\omega )=
-j_\mu^*( \inn_{\tilde{\xi}^{M_1}}\d h_1)=0\; .$$ Therefore,
$\Gamma_1$ is tangent to ${\cal J}^{-1}(\mu )$. Moreover, if
$\Gamma_\mu\in\vf ({\cal J}^{-1}(\mu ))$ is a vector field such
that $j_{\mu *}\Gamma_\mu =\Gamma_1\vert_{{\cal J}^{-1}(\mu )}$,
then
 \[
\inn_{\Gamma_\mu}\omega_\mu -\d h_\mu =
j_\mu^*(\inn_{\Gamma_1}\omega_1 -\d h_1)=0\; , \] so the dynamical
equation
\begin{equation}
\inn_{\Gamma_\mu}\omega_\mu -\d h_\mu =0 \label{ecotra}
\end{equation} is compatible and its solutions are 
$\Gamma_\mu+\ker\,\omega_\mu$.
 \hfill $\Box$

The last step is to obtain the orbit space 
$({\cal J}^{-1}(\mu)/G_\mu ,\hat\omega )$ (see Theorem \ref{Marsden-Weinstein}).
Consider the presymplectic Hamiltonian system 
$({\cal J}^{-1}(\mu),\omega_\mu ,h_\mu)$,
 and the canonical projection
 $\pi_\mu\colon{\cal J}^{-1}(\mu )\to{\cal J}^{-1}(\mu )/\ker\,\omega_\mu$.
As $(M_1,\omega_1)$ is symplectic, 
${\cal J}^{-1}(\mu)/G_\mu={\cal J}^{-1}(\mu)/\ker\,\omega_\mu$.
Moreover, by the Marsden-Weinstein theorem, a symplectic form
$\hat\omega\in\df^2({\cal J}^{-1}(\mu)/\ker\,\omega_\mu)$
exists such that $\omega_\mu=\pi_\mu^*\hat\omega$. Then:

\begin{prop}
The function $h_\mu$ and the vector field 
$\Gamma_\mu\in\vf ({\cal J}^{-1}(\mu ))$ satisfying (\ref{ecotra}) are
$\pi_\mu$-projectable, and 
$({\cal J}^{-1}(\mu)/\ker\,\omega_\mu ,\hat\omega ,\hat{h})$ is a symplectic
Hamiltonian system, where $\pi_\mu^*\hat{h}=h_\mu$.
\end{prop}
\proof
 In fact, $\Lie_{\tilde{\xi}^{M_1}_\mu}
h_\mu =0$, for every $\tilde{\xi}^{M_1}_\mu\in \tilde{\bf
g}^{M_1}_\mu\subset\tilde{\bf g}_{M_1}$, since $h_1$ is
$G$-invariant and then $h_\mu$ is $G_\mu$-invariant. Furthermore,
for every $\tilde{\xi}^{M_1}_\mu\in\tilde{\bf g}^{M_1}_\mu$, since
$\omega_\mu$ and $h_\mu$ are $G_\mu$-invariant, we have
\[
\inn_{[\tilde{\xi}^{M_1}_\mu,\Gamma_\mu]}\omega_\mu =
\Lie_{\tilde{\xi}^{M_1}_\mu} \inn_{\Gamma_\mu}\omega_\mu -
\inn_{\Gamma_\mu}\Lie_{\tilde{\xi}^{M_1}_\mu}\omega_\mu =
\Lie_{\tilde{\xi}^{M_1}_\mu}\d h_\mu =0\; ,
\]
and then $[\tilde{\xi}^{M_1}_\mu,\Gamma_\mu]\in\ker\,\omega_\mu$.
But, as all the elements of $\ker\,\omega_\mu$ can be expressed as
$Z_\mu=f^i\xi_{\mu_i}$, then we also have that
$[Z_\mu,\Gamma_\mu]\in\ker\,\omega_\mu$, for every
$Z_\mu\in\ker\,\omega_\mu$. Therefore, $\Gamma_\mu$ is
$\pi_\mu$-projectable.

Finally, since $\tilde{\bf g}^{M_1}_{\mu_x}=\ker\,\omega_{\mu_x}$, 
for every $x\in{\cal J}^{-1}(\mu)$, then 
${\cal J}^{-1}(\mu)/G_\mu={\cal J}^{-1}(\mu)/\ker\,\omega_\mu$. 
As a consequence, 
$({\cal J}^{-1}(\mu )/\ker\,\omega_\mu,\hat\omega)$ is a symplectic
manifold. Hence, from (\ref{ecotra}), 
$({\cal J}^{-1}(\mu)/\ker\,\omega_\mu ,\hat\omega ,\hat{h})$ is a symplectic
Hamiltonian system and 
\beq 
\inn_{\hat \Gamma}\hat\omega -\d\hat{h}=0 
\label{ecuac} 
\eeq 
where $\pi_{\mu*}\Gamma_\mu=\hat \Gamma$.
 \hfill $\Box$

\subsection{The singular case: symmetries and reduction}

As already pointed out, the concept of group of symmetries can be
extended in a natural way to the case of singular optimal control
problems. Given the presymplectic dynamical system $(M,\omega,H)$,
if the optimal control problem is singular, then when we apply
the constraint algorithm described in Section 3, a submanifold
$M_1$ is obtained. However, the 1-form $\omega_1=j_1^*\omega$ is now
presymplectic and in the best of cases there will exist a
family of vector fields satisfying the dynamical equation
(\ref{conmo}) in the points of $M_1$ and tangent to $M_1$.
Otherwise, we apply the next steps in the constraint algorithm in
order to obtain a final constraint submanifold $M_f$ in which
there exist vector fields $\Gamma\in\vf(M)$ tangent to $M_f$ such
that
\[
(\inn_\Gamma\omega-\dif H)|_{M_f}=0\; . \]

If $j_f\colon M_f\hookrightarrow M$ denotes the embedding, let us
consider the presymplectic dynamical system $(M_f,\omega_f,h_f)$,
where $\omega_f=j_f^*\omega$ and $h_f=j_f^*H$. If the 2-form
$\omega_f$ is nondegenerate, then the study of symmetries and the
geometric reduction of the singular problem proceeds in a similar
way to the regular case, just by replacing, in Definition \ref{sym1},
$(M_1,\omega_1,h)$ with the data $(M_f,\omega_f,h_f)$.
It remains to consider the case when the compatible dynamical
system $(M_f,\omega_f,h_f)$ is presymplectic. In
this case we need to apply the generalization of the Marsden-Weinstein
reduction theory to the case of presymplectic manifolds
(see the Appendix \ref{algpm}), and replacing once again $(M_1,\omega_1,h)$
by $(M_f,\omega_f,h_f)$ in Definition \ref{sym1}.
This action is denoted $\Phi_f$.
In both cases the problem is given by a group $G$ acting on $M$,
and leaving $M_f$ invariant.

If ${\cal J}_f$ is the momentum map associated with the
presymplectic action $\Phi_f$ and $\mu\in{\bf g}^*$ is a weakly
regular value, then the submanifold ${\cal J}^{-1}_f(\mu )$ of
$M_f$, the form $\omega_\mu =j_\mu^*\omega_f$ and the function
$h_\mu =j_\mu^*h_f$ make a compatible presymplectic Hamiltonian
system $({\cal J}^{-1}_f(\mu ),\omega_\mu ,h_\mu)$. The proof is
similar to the regular case.

Let us denote by ${\bf \tilde{g}_{M_f}}$ the set of fundamental
vector fields on $M_f$ with respect to the action $\Phi_f$, and
let $\Gamma_f$ be a solution to the dynamical system
$\inn_{\Gamma_f}\omega_f-\dif h_f=0$. For every constraint $\zeta$
defining ${\cal J}^{-1}_f(\mu )$, if $\d\zeta
=\inn_{\tilde{\xi}^{M_f}}\omega_f$, with
$\tilde{\xi}^{M_f}\in\tilde{\bf g}_{M_f}$, then
$j_\mu^*\Gamma_f(\zeta )=0$. That is, the vector fields $\Gamma_f$
are tangent to ${\cal J}^{-1}_f(\mu )$. Moreover, if
$\Gamma_\mu\in\vf ({\cal J}^{-1}_f(\mu ))$ is a vector field such
that $j_{\mu *}\Gamma_\mu =\Gamma_f\vert_{{\cal J}^{-1}_f(\mu )}$,
then the dynamical equation $\inn_{\Gamma_\mu}\omega_\mu -\d h_\mu
=0$ is compatible and its solutions are $\Gamma_\mu
+\ker\,\omega_\mu$. Finally, the procedure to obtain the orbit
space $({\cal J}^{-1}_f(\mu)/G_\mu,\hat{\omega})$ and the
Hamiltonian function $\hat{h}$ follows the same pattern as in the
previous section.

\section{Symmetries and reduction of optimal control problems:
 the non-autonomous case}

In this section, we extend the previous results to the case of
non-autonomous optimal control problems. After a description of a
geometric formulation of the problem, we analyze the geometric
reduction in the regular case. The extension of these results to the singular context
is similar to that of autonomous problems.

\subsection{Geometric description}

If the optimal control problem is non-autonomous, then the control
equation (\ref{ligadura}) becomes
\begin{equation}\label{ligadurat}
\dot{q}^i(t)=X^i(t,q(t),u(t))
\end{equation}
and the objective functional to minimize is
\begin{equation}\label{funcionalt}
S[\gamma]=\int_{t_1}^{t_2} L(t,q(t),u(t))\dif t\; ,
\end{equation}
where at least one of the ``functions'' $X^i$ ($i=1,\dots,n$) or
$L$ depends explicitly on the time. A necessary condition for the
existence of an optimum is still given by the Pontryagin's
maximum principle (\ref{Pon}-\ref{eqham}), where the Hamiltonian
function is (\ref{hamiltoniano}) (if we include the
time-dependence).

Now we provide a geometric description of such equations when the
maximum of $H$ is not obtained on the boundary of the control set.
A non-autonomous optimal control problem may be given by the
following data: a configuration space, which is the trivial bundle
$\rho_1\colon\Real\times Q\to\Real$ (elements of $Q$ describe the state
variables, and $\Real$ is the time); a fibre bundle
$Id\times\pi\colon \Real\times E\to\Real\times Q$ (which is the identity
in the first factor) whose fibres describe the control variables;
a ``Lagrangian function'' $L\colon \Real\times E\to\Real$; and a vector
field $X$ along the projection $\rho_2\circ(Id\times\pi)$ (where
$\rho_2\colon\Real\times Q\to Q$ denotes the projection onto the
second factor), i.e., $X\colon\Real\times E\to\Tan Q$ is such that
$\tau_Q\circ X=\rho_2\circ(Id\times\pi)$. For sections
$\sigma\colon I\to\Real\times E$ ($I=[t_1,t_2]$) such that
$(Id\times\pi)\circ\sigma$ have fixed end-points, the problem is
to find a section minimizing the action
\[
\int_{t_1}^{t_2} L(\sigma)\dif t
\]
when $\sigma$ satisfies the differential equation
\[
\bar{\rho}_2\circ(Id\times\pi)\circ j^1\sigma = X\circ\sigma\; ,
\]
where $\bar{\rho}_2\colon\Real\times\Tan Q\to\Tan Q$ denotes the
projection onto the second factor. So we have the following
commutative diagramme:
 \[
\begin{array}{ccccc}
 & & I\times \Tan E &
\begin{picture}(80,10)(0,0) \put(8,6){\mbox{$\bar{\rho}_2\circ
(Id\times\Tan\pi)$}} \put(0,3){\vector(1,0){80}}
\end{picture} & \Tan Q
\\
 &
 \begin{picture}(50,50)(0,0)
\put(4,25){\mbox{$j^1\alpha$}} \put(0,0){\vector(1,1){50}}
\end{picture}
&
\begin{picture}(10,50)(0,0)
\put(8,20){\mbox{$Id\times\tau_E$}} \put(3,50){\vector(0,-1){50}}
\end{picture}
&
\begin{picture}(80,50)(0,0)
\put(50,19){\mbox{$X$}} \put(0,0){\vector(3,2){80}}
\end{picture}
&
\begin{picture}(10,50)(0,0)
\put(8,20){\mbox{$\tau_Q$}} \put(3,50){\vector(0,-1){50}}
\end{picture}
\\
I &
\begin{picture}(55,10)(0,0)
\put(25,6){\mbox{$\sigma$}} \put(0,3){\vector(1,0){55}}
\end{picture}
& I\times E &
\begin{picture}(80,10)(0,0)
\put(13,6){\mbox{$\rho_2\circ (Id\times\pi)$}}
\put(0,3){\vector(1,0){80}} \end{picture}
 & Q
\end{array}
 \]

The analog of the presymplectic description of autonomous systems
shown in Section 2 is the following: we take the fiber bundle
$\Real\times\pi^*\Tan^*Q$, which has canonical projections
$Id\times\pi_1\colon \Real\times\pi^*\Tan^*Q \to \Real\times E$ and
$Id\times\pi_2\colon \Real\times\pi^*\Tan^*Q \to \Real\times\Tan^*Q$.
Then the Hamiltonian function (\ref{hamiltoniano}) can be defined
intrinsically as
 \begin{equation}\label{hamgeomt}
 H = \hat{X} - L\; ,
 \end{equation}
 where we identify the Lagrangian function $L\in\Cinfty(R\times E)$
 with its pull-back through $Id\times\pi_1$ and
 $\hat{X}\colon\Real\times\pi^*\Tan^*Q\to\Real$ is the function defined by
 $\hat{X}((t,q,u),\alpha)=\alpha(X(t,q,u))$.

If $\Theta$ and $\Omega$ are the pull-backs to
$\Real\times\pi^*\Tan^*Q$ of the canonical forms in $\Tan^*Q$, let
$\Theta_H=\Theta+H\dif t=p_i\dif q^i +H\dif t$. Then the
solutions to the equations of motion (\ref{eqham}) and
(\ref{eqlig}) (which are obtained as necessary conditions from the
maximum Pontryagin's principle if the control variables are
interior points) are obtained from the integral curves of a vector
field $\Gamma\in\vf (\Real\times\pi^*\Tan^*Q)$ verifying
\begin{equation}\label{conmot}
\inn_\Gamma\Omega_H=0\; ,~~~\inn_\Gamma\dif t=1\; ,
\end{equation}
where $\Omega_H=-\dif\Theta_H$, when we restrict the equations to
the maximal manifold where a solution exists.

 Once again, this is a presymplectic system, and a constraint
 algorithm  similar to the one developed in Section 3 should be
 applied. It is easy to prove that, if we denote by
 $\tilde{M}=\Real\times\pi^*\Tan^*Q$, the maximal submanifold
 $\tilde{M}_1$ where a
 solution to equations (\ref{conmot}) exists is described
 by equation (\ref{eqlig}), i.e., $\tilde{M}_1$ is defined locally by
 $\{\varphi_a=0\}$, where $\varphi_a=\partial H/\partial u^a$. In
 this section we will assume that
 $\det\left(\frac{\partial^2 H}{\partial
 u^a\partial u^b}\right)\neq 0$, in such a way that we can solve locally
 the control variables as functions of the other variables,
 $u^a=\Psi^a(t,q,p)$. Then, the algorithm finishes in the
 first step, and there exists a unique vector field
 $\Gamma$ tangent to $\tilde{M}_1$ satisfying
 (\ref{conmot}) in the points
 of $\tilde{M}_1$ (i.e., the optimal control problem is regular).
 In this case, $\tilde{M}_1$ is locally diffeomorphic to
 $\Real\times\Tan^*Q$. If $\tilde{j}_1\colon\tilde{M}_1 \to
 \tilde{M}$ denotes the embedding, let $h_1=H(t,q,p,\Psi(t,q,p))$
 be the pull-back through $\tilde{j}_1$ of $H$. Then
 $\Omega_{1h}=j_1^*(\Omega_H)=\Omega_1+h_1\dif t$, where
 $\Omega_1=j_1^*\Omega$. If $\dim Q=n$, then
$\dim \tilde{M}_1=2n+1$ and $\Omega_{1h}$ is of maximal rank 2n,
in such a way that the pair $(\Omega_{1h},\dif t)$ is a
cosymplectic structure of $\tilde{M}_1$, since
$\Omega_{1h}\wedge\dif t\neq 0$. If $\Gamma_1\in\tilde{M}_1$ is
the unique vector field solution to the dynamical equations
 \[
 \inn_{\Gamma_1}(\Omega_1+h_1\dif t) = 0~,~~\inn_{\Gamma_1}\dif t =
 1\; ,
 \]
 then the vector field $\Gamma$ is obtained from $\Gamma_1$
 by using the feedback condition $u^a=\Psi^a(t,q,p)$ and
 the boundary conditions.

\subsection{Symmetries and  reduction}

Concerning the study of symmetries, time-dependent optimal control
problems display some particular characteristics which are
worth consideration. Following the ideas in \cite{EMR-99}, if
$G$ is a Lie group, $(\tilde{M} ,\Omega_H)$ is a non-autonomous
optimal control system and $\Phi\colon G\times\tilde{M}\to
\tilde{M}$ is an action of $G$ on $\tilde{M}$, $G$ is said to be a
{\sl group of standard symmetries} of this system if, for every
$g\in G$,
 \begin{enumerate}
\item $\Phi$ leaves $\tilde{M}_1$ invariant, i.e., it induces an action
$\Phi_1\colon G\times \tilde{M}_1\to \tilde{M}_1$;
 \item $(\Phi_1)_g$ preserves the forms $\Omega_1$ and $\d t$
 (it is a cosymplectic action), that is, $$
(\Phi_1)_g^*\Omega_1 =\Omega_1 \quad ;\quad (\Phi_1)_g^*\d t =\d
t\; ; $$
\item
and $(\Phi_1)_g$ preserves the dynamical function $h_1$; i.e.,
$(\Phi_1)_g^*h_1=h_1$.
 \end{enumerate}
 The diffeomorphisms $\Phi_g$ are called {\rm standard symmetries} of
the system.

As an immediate consequence of this definition, if $G$ is a group
of standard symmetries of the non-autonomous system $(\tilde{M}
,\Omega_H)$ then, for every $g\in G$, $(\Phi_1)_g$ preserves the
form $\Omega_{1h}$: $(\Phi_1)_g^*\Omega_{1h} -\Omega_{1h} =
j_1^*(\Phi_g^*\Omega_H-\Omega_H) = 0$. Moreover, $G$ is a group of
standard symmetries of the non-autonomous system above if, and
only if, the following three conditions hold for every $\xi\in{\bf
g}$: $$ {\rm (1)}\qquad \Lie_{\tilde\xi }\Omega_1=0 \quad , \qquad
{\rm (2)}\qquad \Lie_{\tilde\xi }\d t=0 \quad , \qquad{\rm
(3)}\qquad \Lie_{\tilde\xi }h_1=0 $$

At this point, reduction of regular non-autonomous optimal control
problems with symmetry follows a similar pattern to the reduction
made above of autonomous optimal control problems with symmetry.
Actually, singular optimal control problems can be studied by
using similar ideas.

\begin{rem} {\rm We would like to point out that the reduced Hamiltonian
system does not describe, in general, an optimal control problem. To
show this, it is enough to recall that every variational problem
can be written as an optimal control problem by taking $E=\Tan Q$
in the control bundle $\Real\times E\to \Real\times Q$, and considering
as control equations $\dot{q}^i=u^i$ (where $(q^i, u^j)$ denote
the local coordinates in $E=\Tan Q$). However, in \cite{MR-01} it is
shown that, in general, the reduced Hamiltonian system is not a
Lagrangian system. To study when the reduced Hamiltonian system
describes an optimal control problem, an inverse problem should be
solved.}
\end{rem}

\section{Examples}

\subsection{Reduction of regular optimal control problems
invariant by a vector field}

In order to illustrate the above results, we study the case where
the optimal control problem is invariant by a vector field, that
is, there exists a vector field $Z\in\vf(E)$ which is an
infinitesimal symmetry of the regular optimal control problem
$(L,\pi,Q,X,L)$ (see Definition \ref{infisym}). As $Z$ is
$\pi$-projectable, let $Z_0=\pi_*Z\in\vf(Q)$. In local
coordinates,
 \[
 Z=f^i(q)\frac{\partial}{\partial q^i} +
 g^a(q,u)\frac{\partial}{\partial u^a}~~~\hbox{and}~~~
 Z_0 = f^i(q)\frac{\partial}{\partial q^i}\; .
 \]
 We can lift $Z\in\vf(E)$ to a new vector field
 $Z^c\in\vf(M)$ (where $M=\pi^*\Tan^*Q$, as usual)
 whose local expression is
 \[
 Z^c = f^i(q)\frac{\partial}{\partial q^i} -
 p_j\frac{\partial f^j}{\partial q^i}\frac{\partial}{\partial p_i} +
 g^a(q,u)\frac{\partial}{\partial u^a}\; .
 \]
 Let $\varphi_t$, $\Psi_t$ and
 $\Phi_t$ be the flows of
 vector fields $Z_0\in\vf(Q)$, $Z\in\vf(E)$ and
 $Z^c\in\vf(M)$, respectively. Let $\Gamma\in
 \vf(M)$ be the vector field in $M$ constructed by the extension
 of the vector field $\Gamma_1\in\vf(M_1)$ solution to the
 optimal control problem in $M_1$, using feedback condition
 (\ref{feedback}). Then $\Phi_t^*\Gamma$ is also a solution to the
 dynamical system in the sense that its restriction to $M_1$,
 is again a vector field tangent to $M_1$ verifying the dynamical
 equation restricted to $M_1$.

 Let us assume that the vector field $Z$ is complete. Then $Z^c$ is
 also complete, and it induces an action
  \[\begin{array}{ccc}
  \Phi\colon\Real\times M & \longrightarrow & M
  \\
  (t,(q,u,p)) & \mapsto & \Phi_t(q,p,u)
  \end{array} \]
  which restricts to a new action
  \[
  \Phi_1\colon\Real\times M_1\longrightarrow M_1\; .
  \]
  Since $Z$ is an infinitesimal symmetry of the optimal control
  problem, then $\Real$ is a symmetry group of the presymplectic
  dynamical system $(M_1,\omega_1,h_1)$ under this action, and
  $f_Z=\langle\theta,Z^c\rangle$ is a constant of motion ($f_Z$ is
  the comomentum map). Let ${\cal J}$ be the dual momentum map, and
  consider the level set
  \[
  {\cal J}^{-1}(\mu)=\{ (q,p)\in M_1\mid p_if^i(q)=\mu\}\; ,
  \]
  where $\mu\in\Real$. If $Z_0$ is non-vanishing everywhere, then
  ${\cal J}^{-1}(\mu)$ is a submanifold of $M_1$. Moreover, in this case,
  $\Tan_{(q,p)}{\cal J}$ is surjective and therefore every $\mu\in\Real$ is a
  regular value. In general, for an arbitrary $\pi$-projectable
  and complete
  vector field $Z$, $\mu\in\Real$ will be neither regular nor
  almost regular.
  If we assume that $\mu$ is at least almost regular, let us
  consider the dynamical system $({\cal J}^{-1}(\mu),\omega_\mu,h_\mu)$
  where $j_\mu\colon {\cal J}^{-1}(\mu)\hookrightarrow M_1$ denotes
  the embedding, $\omega_\mu=j_\mu^*\omega_1=(j_\mu^*\circ
  j_1^*)\omega = (j_1\circ j_\mu)^*\omega$ and $h_\mu=j_\mu^* h_1
  =(j_\mu^*\circ j_1^*) H = (j_1\circ j_\mu)^*H$. Since
  $\dim {\cal J}^{-1}(\mu) = 2\dim Q -1$, then the dynamical system is
  presymplectic, and the reduction procedure finishes quotienting
  by $\ker\omega_\mu$ ($h_\mu$ is projectable under this
  distribution).

  It is interesting to realize that the momentum map ${\cal J}\colon
  M_1\to\Real$ can be extended to a map ${\bf J}\colon M\to\Real$ whose
  local expression coincides with the local expression of the momentum
  map ${\cal J}$. The level sets are again
  \[
  {\bf J}^{-1}(\mu)=\{ (q,u,p)\in M\mid \langle\theta,Z^c
  \rangle = p_if^i(q)=\mu\}\; .
  \]
  Moreover, ${\bf J}$ is a momentum map, since the action is
  strictly presymplectic.

  Let $({\bf J}^{-1}(\mu),\bar{\omega}_\mu,\bar{H}_\mu)$ be the
  presymplectic dynamical system given by $\bar{\omega}_\mu =
  \bar{j}^*_\mu\omega$ and $\bar{H}_\mu=\bar{j}^*_\mu H$, where
  $\bar{j}_\mu\colon{\bf J}^{-1}(\mu)\hookrightarrow M$
  denotes the embedding (again, we assume that $\mu$ is, at least,
  an almost regular value of the momentum map). The presymplectic
  dynamical system
  $({\bf J}^{-1}(\mu),\bar{\omega}_\mu,\bar{H}_\mu)$ has solution
  in the points of ${\cal J}^{-1}(\mu)$.

\subsection{Shortest paths with bounded curvature}

The following example is a free version of a problem which is studied
from a different point of view
in \cite{Su-95} (see also the quoted references),
and consists in characterizing the shortest $C^1$-curves that
are parametrized by arc length satisfying a curvature bound, and
going from a given  initial position and velocity to a final one.

In this model the configuration space is $Q=\Real^3\times S^2$,
with local coordinates $(x^i,y^i)$ ($i=1,2,3$),
with $\sum_i (y^i)^2=1$. The control space is, in principle,
the closed unit ball $B^3$ in $\Real^3$, and hence the
bundle of controls is $E=\Real^3\times S^2\times B^3$,
with coordinates $(x^i,y^i,u^i)$ ($i=1,2,3$).
The differential equations are
$$
\dot{\bf x}={\bf y}
\quad ; \quad
\dot{\bf y}={\bf y}\times{\bf u}
$$
where ${\bf x}\equiv (x^1,x^2,x^3)$, ${\bf y}\equiv (y^1,y^2,y^3)$,
${\bf u}\equiv (u^1,u^2,u^3)$, and
${\bf y}\times{\bf u}$ denotes the cross product in $\Real^3$.
The Lagrangian function for this problem is  $L=1$.

The presymplectic Hamiltonian description of this system is made
in the manifold $E\times_Q\Tan^*Q$, where we have the coordinates
$(x^i,y^i,u^i;p_i,q_i)$ ($p_i,q_i$ are the conjugate momenta associated
with the position coordinates $x^i,y^i$).
The canonical forms are then
$$
\theta=p_i\d x^i+q_i\d y^i
\quad , \quad
\omega=\d x^i\wedge\d p_i+\d y^i\wedge\d q_i
$$
The Hamiltonian function is
$$
H=\langle{\bf p},{\bf y}\rangle+\langle{\bf q},{\bf y}\times{\bf u}\rangle +1
$$
(where $\langle ,\rangle$ denotes the usual scalar product in $\Real^3$,
arising from the duality between $\Tan Q$ and $\Tan^*Q$).
Observe that $H$ is linear on the controls, and so it is known that
the optimal solutions for the controls are in the boundary of $B^3$;
that is in $S^2$, unless ${\bf y}\times{\bf u}=0$, when every value
of the controls gives an optimal solution. Hence we can take
$E=\Real^3\times S^2\times S^2$.

This system exhibits symmetries which are:
\begin{itemize}
\item
Rigid translations in $\Real^3$,
whose action on $E=Q\times S^2$ is as follows:
for a given ${\bf v}\in\Real^3$, if $\tau_v\colon\Real^3\to\Real^3$
is the translation ${\bf x}\mapsto{\bf x}+{\bf v}$, we have that
$\tau_v({\bf x},{\bf y},{\bf u})=({\bf x}+{\bf v},{\bf y},{\bf u})$.
\item
Rotations on $\Real^3$ which act on $E$ in the following way:
for a given rotation $R\in SO(3)$, we have that
$R({\bf x},{\bf y},{\bf u})=(R{\bf x},R{\bf y},R{\bf u})$.
\end{itemize}
That is, the group of symmetries is $G=\Real^3\times SO(3)$.
The infinitesimal generators are the
following vector fields in $E$
\beann
\tilde\xi_i &=&\derpar{}{x^i} \qquad (i=1,2,3)
\\
\tilde\xi_4 &=&
x^1\derpar{}{x^2}-x^2\derpar{}{x^1}+y^1\derpar{}{y^2}-y^2\derpar{}{y^1}+
u^1\derpar{}{u^2}-u^2\derpar{}{u^1}
\\
\tilde\xi_5 &=&
x^2\derpar{}{x^3}-x^3\derpar{}{x^2}+y^2\derpar{}{y^3}-y^3\derpar{}{y^2}+
u^2\derpar{}{u^3}-u^3\derpar{}{u^2}
\\
\tilde\xi_6 &=&x^3\derpar{}{x^1}-x^1\derpar{}{x^3}+y^3\derpar{}{y^1}-y^1\derpar{}{y^3}+
u^3\derpar{}{u^1}-u^1\derpar{}{u^3}
\eeann
whose canonical liftings to $E\times_Q\Tan^*Q$ give the following
fundamental vector fields
\beann
\tilde\xi_i^c &=&\derpar{}{x^i} \qquad (i=1,2,3)
\\
\tilde\xi_4^c &=&
x^1\derpar{}{x^2}-x^2\derpar{}{x^1}+y^1\derpar{}{y^2}-y^2\derpar{}{y^1}+
u^1\derpar{}{u^2}-u^2\derpar{}{u^1}+
\\ & &
p^1\derpar{}{p^2}-p^2\derpar{}{p^1}+q^1\derpar{}{q^2}-q^2\derpar{}{q^1}
\\
\tilde \xi_5^c &=&
x^2\derpar{}{x^3}-x^3\derpar{}{x^2}+y^2\derpar{}{y^3}-y^3\derpar{}{y^2}+
u^2\derpar{}{u^3}-u^3\derpar{}{u^2}+
\\ & &
p^2\derpar{}{p^3}-p^3\derpar{}{p^2}+q^2\derpar{}{q^3}-q^3\derpar{}{q^2}
\\
\tilde \xi_6^c &=&
x^3\derpar{}{x^1}-x^1\derpar{}{x^3}+y^3\derpar{}{y^1}-y^1\derpar{}{y^3}+
u^3\derpar{}{u^1}-u^1\derpar{}{u^3}
\\ & &
p^3\derpar{}{p^1}-p^1\derpar{}{p^3}+q^3\derpar{}{q^1}-q^1\derpar{}{q^3}
\eeann
Thus $\{\tilde\xi_1^c,\tilde\xi_2^c,\tilde\xi_3^c,\tilde\xi_4^c,
\tilde\xi_5^c,\tilde\xi_6^c\}$
is a set of generators of $\tilde{\bf g}$, but observe that
${\rm dim}\,\tilde{\bf g}=5$.
The action considered is strongly presymplectic, since
it is an exact action in relation to the 1-form $\theta$.
The presymplectic Hamiltonian functions
$f_{\xi_j}\in\Cinfty(E\times_Q\Tan^*Q)$ ($j=1,\ldots ,6$)
associated with $\tilde\xi_j^c$ are
\beann
f_{\xi_i} &=& p_i \qquad (i=1,2,3)
\\
f_{\xi_4} &=& x^1p_2-x^2p_1+y^1q_2-y^2q_1
\\
f_{\xi_5} &=& x^2p_3-x^3p_2+y^2q_3-y^3q_2
\\
f_{\xi_6} &=& x^3p_1-x^1p_3+y^3q_1-y^1q_3
\eeann
So a momentum map ${\cal J}$ can be defined for this action, and
for every weakly regular value $\mu\equiv (\mu_1,\ldots ,\mu_6)\in{\bf g}^*$,
its level sets ${\cal J}^{-1}(\mu)$ foliate $E\times_Q\Tan^*Q$,
and are defined as submanifolds of $E\times_Q\Tan^*Q$ by the constraints
$f_{\xi_j}=\mu_j$ ($j=1,\ldots ,6$); that is, they are made by the points
where the vectors ${\bf p}$ and ${\bf x}\times{\bf p}+{\bf y}\times{\bf q}$
are constant. Observe that, locally, only 5 of these constraints are functionally
independent and, as
${\rm dim}\,(E\times_Q\Tan^*Q)=12$, then  the submanifolds ${\cal J}^{-1}(\mu)$
are 7-dimensional (and presymplectic).
Locally, each one of them can be described by coordinates $(x^i,u^i,z^k)$,
($i=1,2,3$; $k=1,2$) ,where $z^k$ are coordinates which
can be chosen from the set $(y^i,q_i)$.
Next, the final step of the reduction procedure consists in constructing
the quotient manifolds $({\cal J}^{-1}(\mu)/G_\mu,\hat\Omega_\mu)$
(with natural projections
$\sigma_\mu\colon{\cal J}^{-1}(\mu)\to{\cal J}^{-1}(\mu)/G_\mu$).
First notice that all the fundamental vector fields
$\tilde\xi^j$ are tangent to the submanifolds ${\cal J}^{-1}(\mu)$,
hence the isotropy group is $G_\mu=G$, and
the quotient manifolds ${\cal J}^{-1}(\mu)/G_\mu$ are 2-dimensional.
They are described locally by coordinates $(w^k)$ ($k=1,2$),
such that $\sigma_\mu^*w^k$ are functions of the coordinates $(u^i,z^k)$.

As a particular case, we can analyze when
$\mu=0$. Then the constraints defining the submanifold ${\cal J}^{-1}(0)$ are
$$
p_i=0 \qquad (i=1,2,3) \quad ; \quad y^1q_2-y^2q_1=0 \quad ,\quad y^2q_3-y^3q_2=0
\quad , \quad y^3q_1-y^1q_3=0
$$
that is, ${\bf p}=0$, and ${\bf y}\times{\bf q}=0$ (i.e.; ${\bf y}=\lambda{\bf q}$,
with $\lambda\in\Real)$.
Observe that, if
$j_0\colon{\cal J}^{-1}(0)\hookrightarrow E\times_Q\Tan^*Q$
denotes the natural embedding, then
$$
H_0:=j_0^*H=1 \quad , \quad \omega_0:=j_0^*\omega=0
$$
Therefore, in the quotient manifolds ${\cal J}^{-1}(0)/G_0$ we have
$$
\hat H_0=1 \quad , \quad \hat\omega_0=0
$$
where $\hat H_0\in\Cinfty ({\cal J}^{-1}(0)/G_0)$
and $\hat\omega_0\in\df^2({\cal J}^{-1}(0)/G_0)$
are such that $\sigma_0^*\hat H_0=H_0$ and $\sigma_0^*\hat\omega_0=\omega_0$.
Hence, the dynamical equation in ${\cal J}^{-1}(0)/G_0$
has as solutions all the vector fields
$\hat X_0\in\vf({\cal J}^{-1}(0)/G_0)$,
whose local expressions are
$$
\hat X_0=\hat F_k(w)\derpar{}{w^k}
\qquad (\hat F_k\in\Cinfty ({\cal J}^{-1}(0)/G_0))
$$
This result agrees with the analysis made in \cite{Su-95}.

\section*{Appendix: Actions of Lie groups on presymplectic manifolds and reduction}
\protect\label{algpm}

 (This appendix is a review of the results given in \cite{EMR-99}).

Given the presymplectic manifold $(P,\Omega)$, a vector field
$Y\in\vf (P)$ is said to be a Hamiltonian vector field if
$\inn_Y\Omega$ is an exact 1-form; that is, there exists
$f_Y\in\Cinfty (P)$ (the Hamiltonian function) such that
 $\inn_Y\Omega =\dif f_Y$.
We denote by $\vf_h(P)$ the set of Hamiltonian vector fields in
$P$. A function $f\in\Cinfty (P)$ is a Hamiltonian function if
there exists a vector field $Y_f\in\vf (P)$ such that the above equation
holds. We denote by $\Cinfty_h(P)$ the set of Hamiltonian
functions in $P$. A vector field $Y\in\vf (P)$ is said to be a locally Hamiltonian
vector field if $\inn_Y\Omega$ is a closed 1-form. We denote by
$\vf_{lh}(P)$ the set of locally Hamiltonian vector fields in $P$.
Clearly, $\vf_h(P)\subset\vf_{lh}(P)$. Furthermore, $Y\in\vf_{lh}(P)$ if
and only if $\Lie_Y\Omega =0$. For every $Y\in\vf_{lh}(P)$ and
$Z\in\ker\,\Omega$, we have that $[Y,Z]\in\ker\,\Omega$.

Now, let $\Phi\colon P\to P$ be a diffeomorphism.  $\Phi$ is said
to be a {\sl canonical transformation} for the presymplectic
manifold $(P,\Omega)$ if $\Phi^*\Omega =\Omega$. In a similar way,
if $Y\in\vf(P)$ is a vector field such that its flow $\Phi_t$
satisfies $\Phi_t^*\Omega = \Omega$, then $Y$ is said to be an
{\sl infinitesimal canonical transformation} of the presymplectic
manifold. It is clear that $\Phi_t^*\Omega = \Omega$ if, and only
if, $\Lie_Y\Omega=0$ and, hence, $Y$ is an infinitesimal canonical
transformation if, and only if, it is a locally Hamiltonian vector
field.

Let $G$ be a connected Lie group, ${\bf g}$ its Lie algebra and
$\Phi \colon G\times P \to P$ a presymplectic action of $G$ on
$(P,\Omega)$; that is, $\Phi_g^*\Omega = \Omega$, for every $g\in
G$. As a consequence, the fundamental vector fields
$\tilde\xi\in\vf (P)$, associated with $\xi\in{\bf g}$ by $\Phi$,
are locally Hamiltonian vector fields, $\tilde\xi\in\vf_{lh}(P)$
(conversely, if for every $\xi\in{\bf g}$ we have that
$\tilde\xi\in\vf_{lh}(P)$, then $\Phi$ is a presymplectic action
of $G$ on $P$). Therefore, for every $\xi\in {\bf g}$,
$\Lie_{\tilde\xi}\Omega = 0$. We denote by $\tilde{\bf g}$ the set
of fundamental vector fields. When $\tilde{\bf
g}\subseteq\vf_h(P)$, the action $\Phi$ is said to be {\sl
strongly presymplectic} or {\sl Hamiltonian}. Otherwise, $\Phi$ is
called {\sl weakly presymplectic} or {\sl locally Hamiltonian}. In
particular, if $(P,\Omega )$ is an exact presymplectic manifold,
$\Omega=-\dif\Theta$, and the action $\Phi$ is exact (that is,
$\Phi_g^*\Theta = \Theta$, for every $g\in G$), then $\Phi$ is
strongly presymplectic and the fundamental vector fields are
Hamiltonian, with associated Hamiltonian functions
$f_\xi=\inn_{\tilde\xi}\Theta$.

Given a presymplectic action $\Phi$ of a connected Lie group $G$
on the presymplectic manifold $(P,\Omega )$, the {\sl comomentum
map} associated with $\Phi$, \cite{So-69}, is a map (if it exists)
\[
\begin{array}{ccccc}
{\cal J}^*&\colon&{\bf g}&\longrightarrow&\Cinfty_h(P)
\\
& &\xi&\mapsto&f_\xi
\end{array}
\]
where, if $\xi\in{\bf g}$, and $\tilde{\xi}$ is its associated
fundamental vector field, then $f_\xi$ is the function such that
$\inn_{\tilde\xi}\Omega =\dif f_\xi$. The {\sl momentum map}
associated with $\Phi$ is the dual map of the comomentum map; in
other words, it is a map ${\cal J}\colon P\to{\bf g}^*$ such that,
for every $\xi\in{\bf g}$ and $x\in P$,
 $$
 ({\cal J}(x))(\xi):={\cal J}^*(\xi)(x)=f_\xi(x)\; .
 $$

 From the definitions, it follows that both the
comomentum and momentum maps exist if, and only if, the
presymplectic action $\Phi$ on $(P,\Omega)$ is strongly
presymplectic (in particular, if $\Phi$ is exact then both
mappings exist).
In general, a comomentum map is not a Lie algebra homomorphism. An
action $\Phi$ is said to be {\sl Poissonian} or {\sl strongly
Hamiltonian} if there exists a comomentum map which is a Lie
algebra homomorphism. Once again, if the action $\Phi$ is exact,
then $\Phi$ is Poissonian and the comomentum map is given by
${\cal J}^*(\xi)=\inn_{\tilde\xi}\Theta$,
for every $\xi\in{\bf g}$.

Let us assume that $\Phi$ is strongly presymplectic. If ${\cal J}$
is the associated momentum map, then an element $\mu \in {\bf
g}^*$ is a {\sl weakly regular value} of ${\cal J}$ if
${\cal J}^{-1}(\mu )$ is a submanifold of $P$, and
$\Tan_x({\cal J}^{-1}(\mu ))=\ker\,\Tan_x{\cal J}$,
for every $x\in{\cal J}^{-1}(\mu )$.
Moreover, if $\Tan_x{\cal J}$ is surjective for every $x\in{\cal
J}^{-1}(\mu)$, then $\mu$ is said to be a {\sl regular value}. In
this paper. Here, every action $\Phi$ is assumed to be {\sl Poissonian, free} and
{\sl proper}, and $\mu\in{\bf g}^*$ is a {\sl weakly
regular value} of ${\cal J}$. We denote by
 $j_\mu\colon{\cal J}^{-1}(\mu )\hookrightarrow P$
 the corresponding immersion.

Next, we give a brief description of ${\cal J}^{-1}(\mu)$, for
every almost regular value $\mu\in{\bf g}^*$. If $\{\xi_i\}$ is a
basis of ${\bf g}$ with dual basis $\{\alpha^i\}$ in ${\bf g}^*$,
by writing $\mu=\mu_i\alpha^i$, a simple computation shows that there
exist Hamiltonian functions associated with the fundamental vector
fields $\{\tilde{\xi}_i\}$ such that ${\cal J}^{-1}(\mu)=\{ x\in
P\mid f_{\xi_i}(x)=\mu_i\}$. In particular, if $\xi\in{\bf g}$ is
such that $\tilde{\xi}\in\ker\,\Omega$, the Hamiltonian functions
can be taken to be equal to zero and, in this case, $\langle \mu,
\xi\rangle=0$.

The connected components of the level sets of the momentum map
${\cal J}$ can be also obtained as the connected maximal integral
submanifolds of the Pfaff system $\inn_{\tilde\xi}\Omega =0$, for
$\tilde\xi\in\tilde{\bf g}$. Therefore, if $x\in{\cal
J}^{-1}(\mu)$, then $\Tan_x{\cal J}^{-1}(\mu)=\tilde{\bf
g}_x^\bot$. As a consequence, since
$\ker\,\Omega_x\subset\tilde{\bf g}_x^\bot$, then
$\ker\,\Omega\subset\underline{\vf({\cal J}^{-1}(\mu))}$ (where
$\underline{\vf ({\cal J}^{-1}(\mu ))}$ denotes the set of vector
fields of $\vf (P)$ which are tangent to ${\cal J}^{-1}(\mu )$).
If the action is exact, then $f_\xi=-\inn_{\tilde\xi}\Theta$ and
the Pfaff system $\inn_{\tilde\xi}\Omega =0$ can be
expressed as $\dif(\inn_{\tilde\xi}\Theta) =0$.

Let $G_{\mu}$ be the isotropy group of $\mu$ for the coadjoint
action of $G$ on ${\bf g}^*$. Then $G_\mu$ is the maximal subgroup
of $G$ which leaves ${\cal J}^{-1}(\mu )$ invariant. So, the
quotient ${\cal J}^{-1}(\mu )/G_\mu$ is well defined and it is
called the {\sl reduced phase space} or the {\sl orbit space} of
${\cal J}^{-1}(\mu )$. The Lie algebra $\tilde{\bf g}_\mu$ of
$G_\mu$ is made of vector fields tangent to ${\cal J}^{-1}(\mu )$,
and we have that $\tilde{\bf g}_\mu =\tilde{\bf
g}\cap\underline{\vf ({\cal J}^{-1}(\mu ))}$.

At this point, we indicate two different possibilities. If
$\tilde{\bf g}\cap\ker\Omega=\{0\}$, then all the fundamental
vector fields give constraints which are not constant functions,
and $\dim{\cal J}^{-1}(\mu)<\dim P$. On the other hand, if
$\tilde{\bf g}\cap\ker\Omega\neq\{0\}$, only the fundamental
vector fields not belonging to $\ker\Omega$ give constraints which
are not constant functions, and $\dim{\cal J}^{-1}(\mu)\leq\dim P$.
Anyway, ${\cal J}^{-1}(\mu)$ inherits a presymplectic structure
$\Omega_\mu :=j_\mu^*\Omega$, whose characteristic distribution is
$\ker\,\Omega_{\mu_x}=\tilde{\bf g}_{\mu_x} +\ker\Omega_x$, for
every $x\in{\cal J}^{-1}(\mu)$.

Finally, the generalization of the Marsden-Weinstein
reduction theorem \cite{MW-rsms} to presymplectic
actions of Lie groups on presymplectic manifolds is:

\begin{theorem}\label{Marsden-Weinstein}
The orbit space ${\cal J}^{-1}(\mu )/G_\mu$ is a differentiable
manifold. If $\sigma\colon{\cal J}^{-1}(\mu )\to{\cal J}^{-1}(\mu
)/G_\mu$ denotes the canonical projection, then there is a closed
2-form $\hat\Omega\in\Omega^2({\cal J}^{-1}(\mu )/G_\mu)$ such
that $\Omega_\mu =\sigma^*\hat\Omega$ (that is, $\Omega_\mu$ is
$\sigma$-projectable), and:
\begin{itemize}
\item
$\hat\Omega$ is symplectic if, and only if, for every $x\in{\cal
J}^{-1}(\mu)$, $\tilde{\bf g}_{\mu_x} =\ker\,\Omega_{\mu_x}$ or,
what is equivalent, $\ker\Omega_x\cap\Tan_x{\cal J}^{-1}(\mu
)\subseteq\tilde{\bf g}_{\mu_x}$.
\item
Otherwise $\hat\Omega$ is presymplectic. In particular, for every
$x\in{\cal J}^{-1}(\mu)$, if $\ker\,\Omega_x\subset\Tan_x{\cal
J}^{-1}(\mu )$ and $\tilde{\bf g}_x\cap\ker\Omega_x =\{ 0\}$, then
${\rm rank}\,\hat\Omega = {\rm rank}\,\Omega$.
\end{itemize}
\label{MWt}
\end{theorem}

\subsection*{Acknowledgments}

We acknowledge the financial support of
{\sl Ministerio de Ciencia y Tecnolog\'\i a},
CICYT PB98-0920 and BFM2002-03493.
We wish to thank Mr. Jeff Palmer for his
assistance in preparing the English version of the manuscript.
We are grateful to the referees, whose suggestions have enabled us to improve
the final version of the work.

\end{document}